\documentclass[sigconf]{acmart}
\renewcommand\footnotetextcopyrightpermission[1]{} %
\usepackage{subcaption}
\usepackage{tikz} 
\usepackage{xcolor}
\usepackage{cleveref}

\AtBeginDocument{%
  }

\usepackage{bm}
\usepackage{enumitem}
\newcommand{\Prob}{\mathbb{P}}

\definecolor{firebrick}{HTML}{B22222} 
\definecolor{clightgray}{rgb}{0.83, 0.83, 0.83}

\begin{document}

\title{
Exploring Next Token Prediction For Optimizing Databases
}

\author{
Yeasir Rayhan and Walid G. Aref
}
\affiliation{
  \institution{Purdue University, West Lafayette, IN, USA}
  \city{}
  \country{}
}
\email{{yrayhan, aref}@purdue.edu}

\renewcommand{\shortauthors}{
Yeasir Rayhan and Walid G. Aref
}

\begin{abstract}
The {\em Next Token Prediction} paradigm (NTP, for short) lies at the forefront of modern large foundational models 
that 
are pre-trained on diverse and large datasets.  These models generalize effectively, and have proven to be very successful in Natural Language Processing (NLP). 
Inspired by the generalization capabilities of Large Language Models (LLMs), we investigate whether the same 
NTP paradigm can 
be applied to DBMS design and optimization tasks. Adopting NTP directly for database optimization is non-trivial due to the fundamental differences between the domains. In this paper, we present a framework, termed Probe and Learn (PoLe), for applying NTP to optimize database systems. 
PoLe leverages Decision Transformers and hardware-generated tokens to effectively incorporate NTP into database systems. 
As a proof of concept, we demonstrate PoLe in the context of the 
index scheduling task over NUMA servers in main-memory database systems.
Preliminary results for 
this scheduling task
demonstrate that adopting NTP and PoLe can improve both performance and generalizability.

\end{abstract}

\maketitle

\section{Introduction}
Hardware and workload are the two dominant factors that have shaped DBMS research over the past 50 years, and are termed the ``Game Changers'' in databases~\cite{natassa2021}. Both factors are rapidly evolving, and are consistently reshaping the design principles and optimization techniques 
of
modern DBMSs. On the hardware front, recent examples include designs and optimizations on GPUs, e.g.,~\cite{BoeschenB22,MageirakosMKCA22,SioulasCKAA19,ChrysogelosKAA19,HeY11,Bress13}, Persistent Memory (PMEM), e.g.,~\cite{Zuo0W18,RenenLK0HODHS18,ZhouAPC21,ArulrajPD15}, 
SSDs, e.g.,~\cite{TsirogiannisHSWG09,ShahHWG08, LeisHK018,HaasL23}, Compute Express Link (CXL), e.g.,~\cite{AhnCLGKJRPMK22,0001LBC24,LernerA24,AhnWMLDBKSRR24} and NUMA servers, e.g.,~\cite{TeubnerM11, AlbutiuKN12,LiPMRL13, BalkesenATO13,LangLANK13, SchuhCD16, PorobicPBTA12,PorobicLTA14,LeisBK014,WagnerK021,PsaroudakisSMSA15,PsaroudakisSMSA16}. 
Recently,
ARM processors have made significant progress in high-performance computing, and are becoming competitive with the x86 processors with high-end servers, e.g., Apple Silicon~\cite{apple-silicon}, Amazon Graviton~\cite{amazon-graviton}, and NVIDIA Grace~\cite{nvidia-cpu}.
On the workload front, the landscape is equally dynamic, and 
has 
taken a drastic turn over the 
recent
years, with the rise of the Large Language Model (LLM, for short) workloads. LLM workloads impose distinct computational and memory demands, further pushing the boundaries of DBMS design and optimization~\cite{abs-2403-05821}.

Another game-changing factor is the heterogeneity of both the hardware and the workloads. On the hardware front, this heterogeneity manifests itself in a variety of compute and memory chips. For example, modern servers are equipped with large cores, small cores, or a mixture of both. Larger cores are preferable for serialized execution, whereas parallelization 
favors
smaller cores. Likewise, nowadays the same Integrated Memory Controllers (IMCs, for short) support both DRAM and PMEM. 
In addition, the heterogeneity leads to variable read and write access latencies across hardware, e.g., in Non Uniform Memory Access (NUMA) servers accessing data that resides in a remote DRAM chip can incur up to $4\times$ more latency than accessing data that resides in the local DRAM chip~\cite{BangMPB20}. 
{Accessing the PMEM can be $3\times$ slower than accessing a local DRAM chip~\cite{abs-1903-05714}.} 
On the workload front, this heterogeneity stems from the diverse range of applications that DBMSs must support, including OLTP, OLAP, HTAP, and modern ML workloads. Each of these hardware and workload applications has distinct performance characteristics that require customized optimization strategies. 
Moreover, the resurgence of 
cloud databases 
has taken over traditional on-site DBMSs. Amazon Elastic Compute Cloud (EC2)~\cite{amazon-ec2} provides as many as \verb|750| instance choices equipped with a diverse range of 
processors, memory, storage, and networks 
hardware to fit different workload applications. 

To address these challenges, 
DBMS engines 
need to 
generalize across heterogeneous hardware and workload applications without sacrificing performance. We need a complete paradigm shift in how DBMSs are designed; one that embraces  hardware and workload heterogeneity to gain the best performance, while  being able to generalize and adapt across new hardware architectures and workload applications without manual intervention. Rather than constantly reinventing the wheel  to keep pace with the ever-changing hardware and workload landscape, the goal is to distill decades of DBMS knowledge for diverse workloads and hardware into a compact, transferable form that is both generalizable and adaptable.

Large Language Models (LLMs)~\cite{BrownMRSKDNSSAA20,Radford2019LanguageMA,Radford2018ImprovingLU,abs-2302-13971,abs-2303-08774,abs-2401-02954} have proven to be very successful in  generalization tasks across multiple fields, including but not limited to, Natural Language Processing (NLP), Computer Vision, and Robotics. Next Token Prediction (NTP) lies at the core of LLMs. It is a foundational concept in modern AI that involves training models to predict the next element in a sequence based on the preceding context. By mastering NTP, LLMs are responsible for a major share of all  AI breakthroughs in many tasks,
e.g., 
machine translation, text summarization, and code generation.
It is only natural to question how this simple yet powerful NTP paradigm can be applied to DBMS design and optimizations, and whether the NTP Paradigm can help in building data systems 
that are 
capable of generalizing across diverse hardware and workload applications while maintaining high performance.

In this paper, we examine the feasibility of applying the Next Token Prediction paradigm (NTP) in the context of database systems. 
We address the challenges associated with adopting this paradigm for database systems, and identify two building blocks, 
namely,
Decision Transformers (DT, for short) and a new type of tokens that we term, DB-tokens. 
DB-tokens are 
hardware profiles that are essential in translating NTP to DBMS optimizations
. We propose a general framework, Probe and Learn (PoLe, for short) for database systems to effectively incorporate the NTP paradigm
, and evaluate its effectiveness through a case-study on the main-memory index scheduling task. Preliminary results on the main-memory index scheduling demonstrate that the PoLe framework can learn novel scheduling techniques improving the index performance in query throughput 
by 
up to 3$\times$
. 

The rest of this paper proceeds as follows. Section~\ref{sec:ntp} discusses the Next Token Prediction paradigm, and how it is non-trivial to translate the Next Token Paradigm to DBMSs. Section~\ref{sec:bb} presents the core building blocks to enable the Next Token Prediction paradigm for database systems. Section~\ref{sec:pole} introduces the PoLe framework. 
To demonstrate the effectiveness of PoLe, in Sections~\ref{sec-case-study} and~\ref{sec:eval}, we present 
a case study of applying PoLe in the context
of scheduling indexing tasks into NUMA servers and some preliminary results as a proof-of-concept.
Finally, Section~\ref{sec:conclusion} concludes the 
paper.


\section{Database Systems And The Next Token Prediction Paradigm}
\label{sec:ntp}
\begin{figure}[htbp]
    \setlength{\belowcaptionskip}{-9.0pt}
    \centering
    \includegraphics[width=0.9\linewidth]{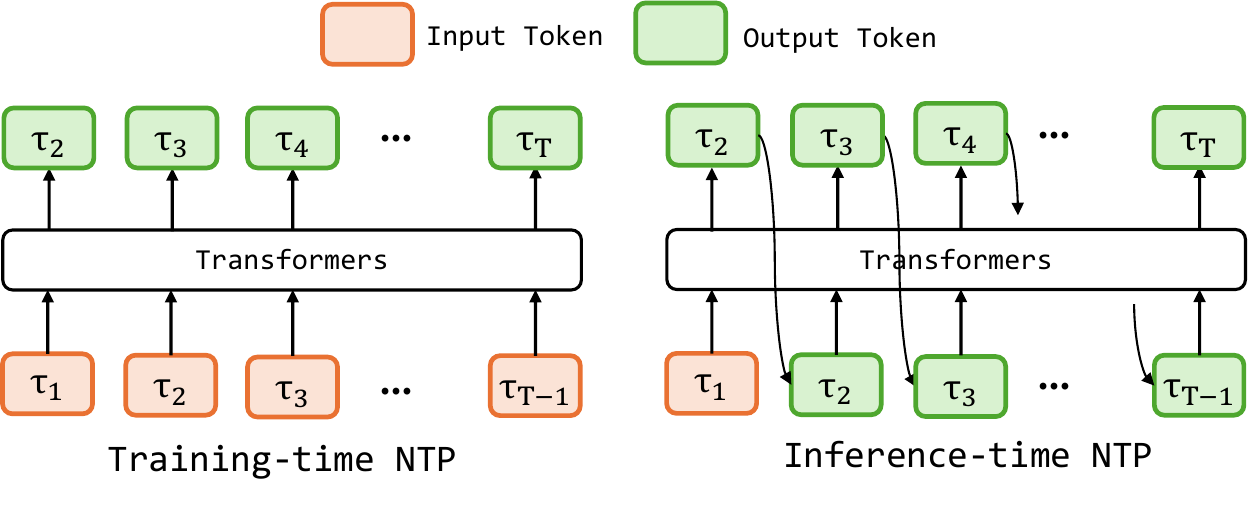}
    \caption{The training and the inference phases of NTP.}
    \label{fig:ntp}
\end{figure}

\noindent\textbf{Next Token Prediction:} At a high level, the next token prediction (NTP, for short) models the probability of the next token in a given sequence based on the past tokens. In Natural Language Processing (NLP), each token is a discrete unit, i.e., a word, a sub-word, or a character. For any sequence $\bm{\tau}$, let $\tau_i$ denote the $i$-th token, and let $\bm{\tau_{<i}}$ denote the $i-1$ tokens preceding $\tau_i$. Then, the goal of NTP is to estimate the probability distribution of $\tau_i$, given $\bm{\tau_{<i}}$, i.e., $\Prob(\tau_i|\bm{\tau_{<i}})$. There are two different phases of NTP, i.e., training-time NTP via self-supervision, and inference-time NTP via autoregression. {Refer to Figure~\ref{fig:ntp} for an illustration of the NTP paradigm during both these phases.}
\begin{itemize}[leftmargin=*]
\item \textbf{Training-time NTP.} During training, the model is fed the previous $\bm{\tau_{<i}}$ ground truth tokens to predict the next token, $\tau_i$. The model is provided with the next ground truth token $\tau_i$ so that it can maximize the sum of log-probabilities over the entire sequence to better predict $\tau_i$ for discrete actions, i.e.,  
$\mathcal{L}_{\text{total}} = -\sum_{i=1}^T \log \Prob(\tau_i \mid \bm{\tau_{<i}})$. This is termed as \textit{Teacher Forcing}~\cite{WilliamsZ89}. 
Notice that, unlike inference, during training the outputs generated by the models are not fed back into itself. As a result, this avoids propagating any compounding error that may result from any misprediction. 
    \vspace{4pt}
    \item \textbf{Inference-time NTP.} During inference, NTP follows an auto-regressive process, i.e., the output generated by the model is fed back into itself to generate the next token. For example, given $\bm{\tau_{<i}}$ tokens, the model generates the next token $\tau_i$ that is then fed back into the model to generate the next token $\tau_{i+1}$. This process continues until a stopping criteria is met. This token generation process is inherently probabilistic, meaning that, at each step, the model outputs a probability distribution over possible next-tokens $\Prob(\tau\mid\bm{\tau_{<i}})$ using either greedy selection, i.e., $\tau_i = \arg\max_{\tau} \Prob(\tau \mid \bm{\tau_{<i}})$, or sampling, i.e., $\tau_i \sim \Prob(\tau \mid \bm{\tau_{<i}})$.
\end{itemize}

\vspace{4pt}
\noindent\textbf{Translating The NTP Paradigm into Database Systems: } 
Many database systems problems can be framed as an NTP task. 
The key question is whether adopting this paradigm effectively aligns with the original objective of the task. For example, consider the Join Order Selection problem (JOS, for short)  that involves 4 tables, i.e., $\mathtt{A \bowtie B \bowtie C \bowtie D}$. Framing 
JOS as an NTP 
task translates 
into
predicting the next table to join. However, this does not align with the objective of JOS that is to determine the join order that minimizes query execution time, rather than simply determining any join order. One of the reasons behind this 
discrepancy is the mismatch in objectives between the domain of NLP and 
that of 
DBMSs. DBMS optimization tasks are often goal-oriented, and are geared towards improving query performance, scalability, and resource utilization. {In contrast, in the era of modern LLMs, NLP tasks often involve generating sequences without explicitly optimizing for any particular objective. This is particularly true during the pre-training phase of LLMs~\cite{Radford2018ImprovingLU}, where the goal is to compress a significant amount of world knowledge into the LLM by training on a diverse internet-scale corpus. }

Moreover, the notion of a token in NLP remains fixed for a particular tokenizer. {NLP tasks additionally benefit from the inherent grammatical rules and the contextual constraints that are consistent throughout the NLP tokens. For example, consider the following sentence: ``SIGMOD'25 is going to be held in Germany''. The verb follows the subject, reflecting syntactic regularity. Contextually, the mention of Germany also implicitly excludes other locations, e.g., USA.
This makes it possible for the NLP tasks to adopt the self-supervised NTP paradigm.}

{
In contrast, the notion of a token in a DBMS can be diverse, e.g., in the Join Order Selection (JOS), the tables can serve as the tokens, whereas in 
query scheduling decision problems, the queries themselves can serve as the tokens. Both of these entities are fundamentally different. Even in the Join Order Selection problem, two different database instances often have distinct sets of tables, making it harder to learn meaningful representation for the tables, i.e., tokens. 
Due to this irregular nature of the concept of tokens in DBMSs, a join order, e.g., $\mathtt{A \bowtie B \bowtie C \bowtie D}$ does not mean much, i.e., both in terms of grammar and context. Without any additional information about the attributes of the 4 relations, a $\mathtt{A \bowtie B \bowtie C \bowtie D}$, join order does not completely eliminate the possibility of another join order, e.g., $\mathtt{D \bowtie C \bowtie A \bowtie B}$. Even if both $\mathtt{A \bowtie B \bowtie C \bowtie D}$ and $\mathtt{D \bowtie C \bowtie A \bowtie B}$ are valid join orders, the crux of JOS lies in identifying the more optimal join order. The table tokens themselves, i.e., A, B, C, D are not enough to distinguish a better join order. This is exactly 
the reason 
why it is challenging to adopt the self-supervised NTP paradigm as 
is for the DBMS optimization tasks, e.g., JOS, 
because it needs 
to install knowledge about the DBMS optimization landscape into a large neural network.} 
Thus, 
as it stands now,
the Next Token Prediction (NTP) paradigm does not directly translate to 
DBMSs.
In the next section, we demonstrate the necessary building blocks to adopt the NTP paradigm for optimizing database systems. 

\section{The Building Blocks}
\label{sec:bb}
One of the key hindrances 
that blocks
DBMSs from benefiting from the NTP paradigm is the goal-directed nature of its optimization tasks. Besides, the notion of token in DBMSs is versatile, i.e., 
is
context-dependent, and as a whole, lacks 
the
generalization capability. To this end, we need goal-conditioned sequence modeling and a redefinition of the concept of tokenization in the context of DBMSs. 
In this section, we discuss the building blocks in adopting the NTP paradigm for optimizing database systems.

\vspace{7pt}
\noindent\textbf{1. Decision Transformer.} Decision Transformer~\cite{ChenLRLGLASM21} (DT, for short) is a promising framework that provides the support for modeling sequences conditioned on a particular objective. 
DT can abstract sequence modeling as a Reinforcement Learning problem (RL, for short), 
where the sequential decision-making is guided by long-term rewards. It 
ignores 
traditional RL techniques, i.e., learning value functions or policies through iterative updates. Rather, it treats RL as a supervised sequence modeling problem, i.e., an NTP problem. Precisely, a DT is an auto-regressive sequence model that predicts the next action based on the past observations conditioned on a desired reward. By conditioning on a desired reward, DT is able to generate more effective next-token predictions, contrary to the traditional NTP paradigm that operates without a specific goal. This reward-conditioned nature of DT allows it to align its next-token predictions with optimal decision-making strategies. This aligns well with DBMS tasks with a pre-defined objective, 
e.g.,
minimizing query execution time. 

Decision Transformers are built on top of the Transformer architecture~\cite{VaswaniSPUJGKP17}. It represents a sequence $\bm{\tau}$ as a combination of 3 types of input tokens, i.e., reward-to-go (RTG, for short), state, and action tokens: $(\hat{r}_0, s_0, a_0, \hat{r}_1, s_1, a_1, \ldots, \hat{r}_{|\bm{\tau}|}, s_{|\bm{\tau}|}, a_{|\bm{\tau}|})$. 
An
RTG token $\hat{r}_t$ denotes the future cumulative reward expected from a given timestep $t$ onward. The initial RTG $\hat{r}_0$ is equal to the reward of the sequence. At Timestep $t$, DT uses the tokens from the latest $K$ timesteps to predict the next action token $a_t$. 
$K$ is the transformer's context length. Then, DT's goal is to learn a policy $\pi_{\texttt{DT}}$ that estimates the probability distribution of the next action $a_t$, given $\bm{s_{\leq t}}$ and $\bm{\hat{r}_{\leq t}}$, i.e., $\pi_{\texttt{DT}}(a_t | \bm{s_{\leq t}}, \bm{\hat{r}_{\leq t}})$. Analogous to the standard NTP paradigm, both the state and action tokens correspond to word tokens in LLMs.

\vspace{7pt}
\noindent\textbf{Training. } The Decision Transformer (DT) is trained following the Offline RL paradigm, i.e., it learns a policy from a fixed dataset $\mathcal{D_{\texttt{OFF}}}$ without online exploration. The dataset $\mathcal{D_{\texttt{OFF}}}$ comprises a collection of experiences, i.e., state-action transitions, $(\hat{r}_i, s_i, a_i)$, generated by various sources,
e.g.,
heuristic methods, RL agents, or human demonstrations. 
DT does not interact with the environment, and is entirely reliant on this fixed dataset $\mathcal{D_{\texttt{OFF}}}$ to learn the best possible policy. Hence, the quality 
and diversity
of the dataset plays a significant role in determining the performance of DT in learning an optimal policy. 
DT samples a mini-batch of experiences with a sequence length of $K$ from $\mathcal{D_{\texttt{OFF}}}$, and trains itself to predict the next action token using the standard L2-error (cross-entropy) for continuous (discrete) actions.

\vspace{7pt}
\noindent\textbf{Inference. } 
During inference, the model is fed the initial state $s_0$ and the desired reward, i.e., performance $\hat{r}_0$. Then, DT 
generates the next action token, $a_0 = \pi_{\texttt{DT}}(\hat{r}_0, s_0)$. Once 
generated, $a_0$ executes to generate the next state token, $s_1 \sim \Prob(s_1|s_0, a_0)$ and the next reward-to-go token, $\hat{r}_1 = \hat{r}_0 - R(s_0, a_0)$. Then, DT is 
fed $\hat{r}_1, s_1$ to generate the next action token prediction $a_1$. This process repeats until all the action tokens in the sequence are generated.


\begin{figure*}[tbp]
    \setlength{\belowcaptionskip}{-5.0pt}
    \centering
    \includegraphics[width=0.85\linewidth]{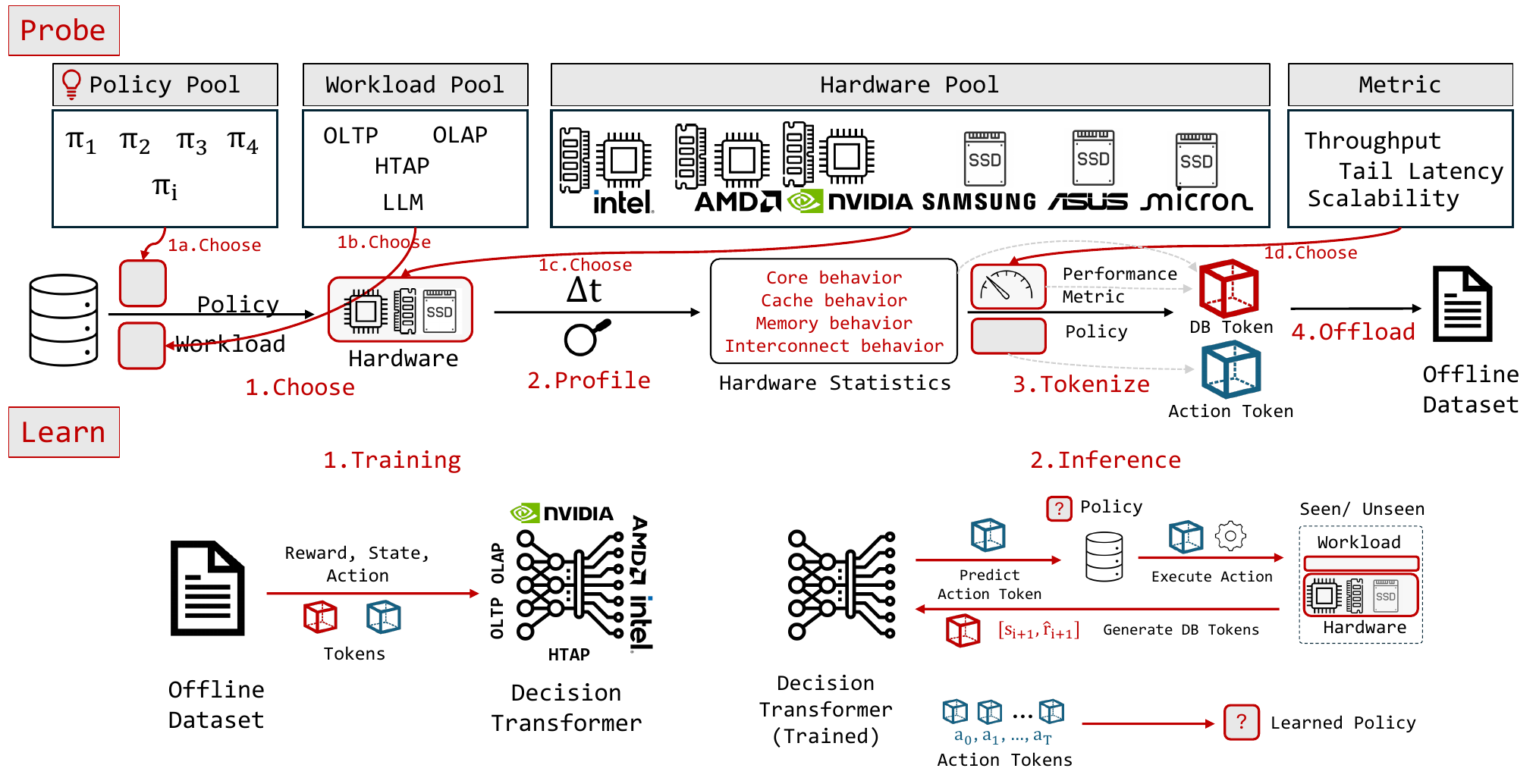}
    \caption{The Probe and Learn (PoLe) framework.}
    \label{fig:pole}
\end{figure*}

\vspace{7pt}
\noindent\textbf{2. Tokenization.}
As discussed 
in Section~\ref{sec:ntp}, the discrete word tokens used in language models do not directly translate to 
database systems. The word tokens in language models serve as linguistic units that remain consistent across all the language tasks, and datasets. In contrast, database systems lack a universal token representation
that
makes it harder to adopt the 
NTP
paradigm 
into DBMSs.

\vspace{7pt}
\noindent\textbf{Database Tokens:} Ideally, the universal token for database systems (DB-token, for short) 
should be able to generalize across new hardware and workload applications. Besides, it should be computationally inexpensive to compute so that it does not affect the day-to-day database tasks. 
Moreover, a DB-token should be able to capture additional context
because in contrast to language models,
these action tokens lack any inherent grammar or structure. These requirements present a significant challenge in identifying an effective internal representation for databases. 
{\bf We propose to use 
as DB-tokens, the 
hardware profiles that are generated 
from hardware Performance Monitoring Units (PMU, for short). }
Hardware profiles are computationally inexpensive to retrieve from the hardware registers, and can generalize across different hardware and workload applications.
Moreover, hardware profiles
can provide accurate hardware context that the DBMS is running on at any given time. 
This makes them an ideal candidate for use as DB-tokens in 
DBMS optimization tasks, as we demonstrate below.

What makes a hardware profile from PMUs stand out in this regard is its capability to provide a real-time update of the hardware and the DBMS components 
at the highest possible granularity. It goes a long way in mimicking the data distribution and capturing the query workload that have proven to be effective in designing efficient learned DBMSs,
e.g., as in~\cite{KraskaABCKLMMN19,PavloAALLMMMPQS17}. For example, 
a small region in the data space that is being queried extensively will result in a lot of cache and memory accesses in the cores that handle this data region. The number of cache accesses may further increase, if the data distribution of that region is particularly dense. Likewise, transactional queries yield less cache and memory accesses compared to the analytical queries that yield a 
large
number of 
executed 
instructions along with higher cache and memory accesses.

Despite the ongoing 
evolution 
in 
the hardware landscape, the 
main
building blocks of the hardware have remained consistent. At its core, there are compute, memory chips, interconnects and storage. Almost all modern processors provide a number of PMU counters to monitor these critical components.
Thus, 
the PMU counters
serve as an excellent fit as 
representative DB-tokens that capture the database system performance across 
varying hardware 
at the  finest possible performance granularity. Hence, DB-tokens are an excellent choice as input tokens in an NTP paradigm for optimizing database systems.

\vspace{7pt}
\noindent\textbf{Action Tokens:} In addition to  DB-tokens, there are action tokens to communicate policies across the DBMS and Machine Learning models. These action tokens represent modular units that make up the policy. The action tokens by themselves hold no inherent meaning. It is the DB-tokens that provide the necessary context, allowing a model to understand and learn the intricacies of DBMS optimization and how specific actions impact performance. 
In the next section, we describe how to connect these two key building blocks, i.e., Decision Transformers and the DB-tokens 
to develop a framework for adopting the NTP paradigm 
into
database systems.


\section{The Probe-and-Learn (PoLe) Framework}
\label{sec:pole}
To effectively adopt the Next Token Prediction (NTP) paradigm in database systems, we propose a 
new
technique, termed Probe-and-Learn (PoLe, for short). For a given DBMS optimization task, this involves probing the hardware during query execution to generate 
DB-tokens. 
These tokens are 
used to train a 
DT to learn an optimal policy for the 
specific
optimization task. 
Refer to Figure~\ref{fig:pole} for an overview of PoLe. 

\begin{figure*}[t]
    \setlength{\belowcaptionskip}{-2.0pt}
    \includegraphics[width=0.9\linewidth]{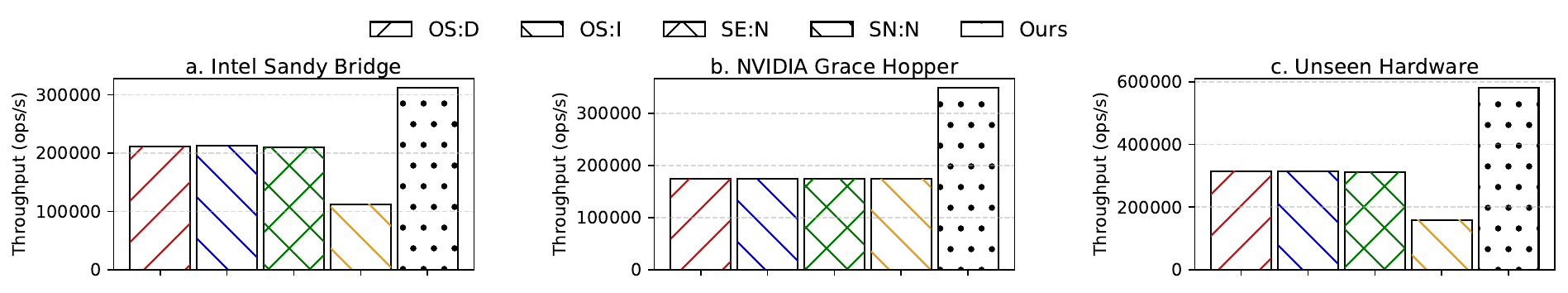}
    \caption{Performance of a main-memory B$^+$-Tree under 
    various
    scheduling policies.}
	\label{fig:exp}
\end{figure*}

\vspace{7pt}
\noindent\textbf{Probe: } The Probe step comprises the following four steps.
\begin{itemize}[leftmargin=*]
    \item \textbf{Choosing.} For a given database optimization task with a set objective, i.e., query throughput, tail latency or scalability, we execute different policies across various hardware configurations and workloads, drawn from a diverse set of policy, hardware, and workload pools. {The policy pool includes both optimal and suboptimal strategies
    that 
    can be randomly sampled or borrowed from existing database literature. The workload pool can comprise a mixture of transactional and analytical workloads. The hardware pool can span servers from different vendors, with varying micro-architectural properties, NUMA configurations, memory technologies, and other distinct characteristics, all of which can significantly impact the performance and behavior of the DBMS.}
    \vspace{2pt}    
    \item \textbf{Profiling.} During query execution, we periodically profile the hardware to capture the behavior of crucial hardware components using hardware statistics, e.g., the number of instructions executed, L1-d, L1-i, L2, Last Level Cache (LLC, for short) {accesses and misses, branch mispredictions, the number of local and remote memory accesses and misses, 
    etc.
    Depending on the DBMS optimization task, the granularity of profiling can vary. For example, for scheduling, profiling can be performed at the query level. In contrast, the join order selection and the data partitioning task requires profiling at the level of query operators and data partitions, respectively. 
    Similarly,
    different optimization tasks can focus on different hardware statistics. For example, NUMA-aware optimizations 
    can
    focus on local and remote memory accesses to capture the data locality. In contrast, the join order selection task can benefit from 
    the write statistics as they reflect the size of the materialized intermediate results.}
    \vspace{2pt}    
    \item \textbf{Tokenization.} Alongside the hardware profiles, we monitor the desired performance metric of the database system, {e.g., the average query throughput, the average tail latency, scalability,
    etc.} These hardware profiles along with the performance metric serve as the DB-tokens. These DB-tokens are analogous to the state and RTG-tokens in DT. Moreover, 
    we 
    tokenize the DBMS policies 
    to generate action tokens. 
    \vspace{2pt}    
    \item \textbf{Offloading.} 
    Periodically, we offload the action tokens along with the associated DB-tokens to an offline dataset {to minimize cache and memory pollution, incurred by this additional data. The offline dataset is 
    provided to
    the DT during the learning phase.}
\end{itemize}

\vspace{3pt}
\noindent\textbf{Learn: } The Learn step comprises the following two steps.
\begin{itemize}[leftmargin=*]
\item \textbf{Training.} We train a Decision Transformer (DT) on the collected offline dataset 
in a supervised manner 
following the Teacher Forcing strategy~\cite{WilliamsZ89}. 
As the DT is trained on diverse policies across varying hardware and workload configurations, it builds a mental model of the performance landscape of the optimization task 
at
hand. It gains insights into the inherent tradeoffs and the performance improvements across different strategies, hardware and workload via the interaction of action and DB-tokens, that are present in the offline dataset. 
\vspace{2pt}    
\item \textbf{Inference.} Using the trained DT, 
PoLe infers 
a new policy 
using auto-regression.
The process begins by feeding DT with the initial DB-token, i.e., state and the desired objective of the specified optimization task. In turn, DT generates the next action token which is fed back to the DBMS to generate the subsequent DB-token. The DB-token is then fed back into DT to get the next action token. This iterative process continues until the complete policy is generated. Note that, during the inference process, the DT has no explicit knowledge about the hardware that the DBMS is running on, or the workload that the DBMS is executing. Hence, both the hardware, and/or the workload can be unseen. It only relies on the DB-tokens, and the inherent knowledge it has gained by being trained on a wide range of policies, hardware configurations, and workloads to generate the next action token.  
\end{itemize}

\section{Case Study: Scheduling Indexing Tasks Over NUMA Servers} 
\label{sec-case-study}
To demonstrate the effectiveness of  PoLe, we apply it to the main-memory index scheduling task. Modern Non Uniform Memory Access (NUMA, for short) servers are notoriously complex and heterogeneous. Each server can contain more than 1000 cores~\cite{intel-sierra-forest} spread across multiple processor dies. Data access latencies can vary significantly by up to 4$\times$ depending on the distance between the core that initiates the data request and the DIMM chip where the data resides. Additionally, concurrently running queries can cause interference by congesting on-chip and off-chip interconnects, as well as saturating hardware queues at caches and memory controllers. Thus, the performance of a DBMS index can vary significantly depending on the physical location of the core where the queries are scheduled, and the DRAM chips where the corresponding data resides. 

We formulate main-memory index scheduling as an NTP task and adopt the PoLe framework to solve it. The index is divided into multiple ($\mathcal{I}_c$) index chunks, i.e., a set of index nodes. Given a set of compute cores, the goal is to predict the core on which the queries should be scheduled for the next index chunk. The data placement is implicit, i.e., the data for each chunk resides in the DIMM chip associated with the assigned core. The core ids serve as the action tokens. The clock-cycles, cache, branch misses, local, and remote DRAM accesses along with the query throughput constitute the DB-token. A complete policy includes a sequence of $\mathcal{I}_c$ core ids. The $i$th core in the sequence represents the core assigned to process the corresponding index slice. 

\begin{figure*}[htbp]
    \includegraphics[width=0.7\linewidth]{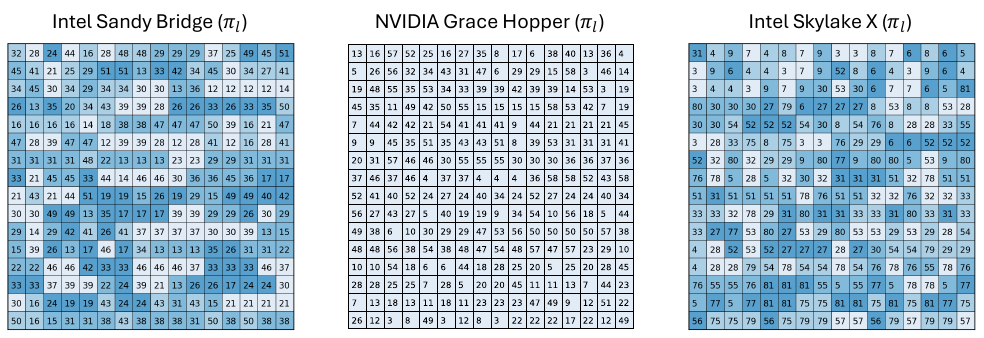}
    \caption{The scheduling policies learned by  PoLe. Each grid corresponds to the learned scheduling policy for the corresponding server for the YCSB-A workload. The top-left grid cell corresponds to the 1st index chunk. The first cell in $i$-th row corresponds to the $\langle i*16+1\rangle$-th index chunk. The corresponding text of a grid cell signifies the 
    hardware core where the query for that particular index chunk is scheduled. Cells sharing the same color indicate that the associated index chunks are scheduled on the same NUMA server.}
	\label{fig:policy}
\end{figure*}
\begin{figure*}[htbp]
    \includegraphics[width=0.8\linewidth]{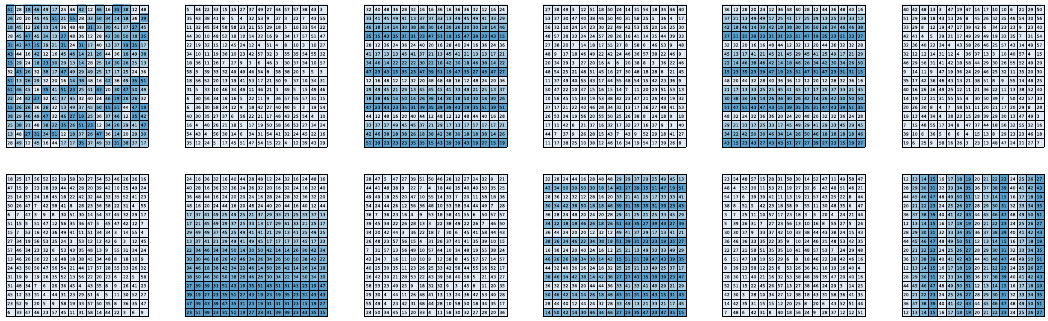}
    \caption{Top performing scheduling policies present in the training dataset for the Intel Sandy Bridge and NVIDIA Grace Hopper servers. Each grid corresponds to a  scheduling policy for the corresponding server. The same scheduling policy can be present multiple times in the training dataset for different workloads. Cells sharing the same color indicate that the associated index chunks are scheduled on the same NUMA server.}
	\label{fig:train_config}
\end{figure*}

\section{Evaluation and Preliminary Results}
\label{sec:eval}
\textbf{Experiment settings: } We evaluate the learned scheduling policy by PoLe against various OS policies and  heuristics. The default OS policy (OS:D) is local allocation, i.e., the OS kernel places an index node on the NUMA node of the requesting CPU. The OS:I policy places the index nodes in an interleaved manner across all the NUMA nodes. 
For both the Shared Everything NUMA (SE:N) and the Shared Nothing~\cite{PorobicPBTA12} (SN) strategies, nearby index slices are clustered together and 
are 
placed in the same NUMA node. For OS:D, OS:I, and SE:N, the OS handles the scheduling. For SN, the scheduling strictly follows the data placement. 

We run a YCSB-like benchmark~\cite{CooperSTRS10} on a main-memory B$^+$-Tree~\cite{WangPLLZKA18,btreecode} on the YCSB-A workload {for 5 minutes}. Initially, the index is loaded with 30M records, and grows up to 200M records as more insertions are performed. Each index record has a 64-bit key and a 64-bit value. For these experiments, the B$^+$-Tree is divided into 256 index chunks. The scheduling policy involves allocating the CPU cores for these 256 index chunks.

{
We evaluate the scheduling policies on the following three  machines:  Intel Sandy Bridge, NVIDIA Grace Hopper, and Intel Skylake X (Unseen Hardware). 
The Intel Sandy Bridge is a 4 socket, 4 NUMA Node, 32 ($\times$2) core machine running Ubuntu 22.04 on a CloudLab~\cite{DuplyakinRMWDES19} d820 node. Each socket in the machine is an 8 ($\times$2) core Intel Xeon E5-4620 processor, with an inclusive 64MB L3 cache. Each physical core in the hardware has a 1MB L1-d, 1MB L1-i, and an 8MB L2 cache. The machine is equipped with 128 GB of DDR3 RAM, and the CPU clock frequency is 2.2 GHz. It does not support Sub-NUMA clustering (SNC) or Cluster On Die (COD) technology. 
The NVIDIA Grace Hopper is a single socket 72 core machine running Ubuntu 22.04 on a CloudLab~\cite{DuplyakinRMWDES19} nvidiagh node. The processor is based on an ARM's Neoverse V2 architecture, where the CPU frequency is 3.1 GHz. The machine is equipped with 
480GB LPDDR5X RAM.
The Skylake-X is a 4 socket, 4 NUMA Node, 92 (×2) core machine running on Ubuntu 22.04. The respective size of L1-d, L1-i, L2, and LLC cache are 3MB, 3MB, 96MB, and 132MB, respectively. The CPU clock frequency is 2.7 GHz.
}

\vspace{7pt}
\noindent\textbf{Preliminary Experimental Results. }The DT in the experiment is trained on tokens generated by executing different YCSB workloads, e.g., YCSB-A, YCSB-C, YCSB-E, on a wide range of hardware 
configurations. 
Figures~\ref{fig:exp}a and~\ref{fig:exp}b illustrate that PoLe can learn better scheduling policies on Intel Sandy Bridge and NVIDIA Grace Hopper, outperforming the baselines by up to 2.78$\times$. For both these experiments, the workload and the hardware have been 
encountered during training, i.e., the offline dataset that the DT has been trained on contains experiences from the same hardware and the same workload. 
Figure~\ref{fig:exp}c illustrates that PoLe can learn a better scheduling policy even for an unseen hardware, i.e., the offline dataset that DT has been trained on does not contain any experience for the particular hardware. It outperforms the baselines by 3$\times$ on an unseen Intel Skylake X 4-Socket 4-NUMA-Node machine. 
Thus, the NTP paradigm not only enhances performance on existing workloads and hardware but also generalizes to optimize index performance in unseen hardware that was not encountered during the training phase.

{
Figure~\ref{fig:policy} gives the scheduling policies learned by the DT for the three corresponding servers. Each grid cell corresponds to an index chunk. The number associated with a grid cell corresponds to the hardware core where the query for that particular index chunk has been scheduled. To ensure that the DT does not merely replicate the best scheduling policies from the offline training dataset, we present the top scheduling policies observed during training in Figure~\ref{fig:train_config}, and it confirms that 
the DT does not replicate the  scheduling policies from the offline training dataset.
}


\vspace{7pt}
\noindent\textbf{Offline Dataset Preparation Cost:} 
{The offline training dataset is generated by executing a diverse set of scheduling policies on a wide range of YCSB workloads on an Intel Sandy Bridge, and an NVIDIA Grace Hopper server. Precisely, the offline dataset 
contains 445 and 49 scheduling episodes, i.e., samples on the two respective servers. Each episode lasts around 5 minutes. Renting an Intel Sandy Bridge-equivalent Haswell server, i.e., a x1.16xlarge~\cite{amazon-ec2-x1} instance on AWS costs approximately \$614 (\$1.380/hr), while renting an NVIDIA Grace Hopper server on Lambda cloud~\cite{nvidia_gh_rent} costs approximately \$73 (\$1.49/hr), totaling an approximate cost of \$687.\footnote{\label{fn:aws_pricing}The AWS pricing info are collected from: \href{https://instances.vantage.sh/?selected=m5a.24xlarge}{https://instances.vantage.sh/}.}
}

\vspace{7pt}
\noindent\textbf{Training Cost: }{We train 
PoLe's
DT on a dual-socket 64 (×2) core machine running Ubuntu 22.04 on a CloudLab~\cite{DuplyakinRMWDES19} r6525 node. The DT comprises approximately 3.37M trainable parameters. It takes approximately 26 CPU hours to train the DT to reach a training accuracy of over 95\%. The inference takes up to 1 CPU minute on the same machine. Renting an equivalent m5a.24xlarge~\cite{amazon-ec2-m5} Amazon AWS instance costs approximately \$32 (\$1.211/hr) to train the DT used in the experiment.}



\section{Concluding Remarks}
\label{sec:conclusion}
In this paper, we lay the foundation for incorporating the Next Token Paradigm in optimizing database systems by proposing a framework that generates hardware tokens to train Decision Transformers in a data-driven framework. 
Whether all or which database optimization tasks can benefit from the proposed PoLe framework as a means to generalize across diverse hardware and workloads 
is yet to be seen.
However, preliminary results on index scheduling over NUMA servers suggest that PoLe has strong potential in improving performance and on adapting to diverse environments. 
Building on these early promising results, a key direction for future work is to conduct an extensive investigation into the database optimization landscape to 
categorize the 
optimization tasks 
that can 
benefit from the Next Token Prediction (NTP) paradigm, and assess to what extent the PoLe framework can provide consistent performance and adaptivity guarantees. 

\balance{}
\bibliographystyle{ACM-Reference-Format}
\bibliography{sample-base}


\begin{thebibliography}{57}


\ifx \showCODEN    \undefined \def \showCODEN     #1{\unskip}     \fi
\ifx \showISBNx    \undefined \def \showISBNx     #1{\unskip}     \fi
\ifx \showISBNxiii \undefined \def \showISBNxiii  #1{\unskip}     \fi
\ifx \showISSN     \undefined \def \showISSN      #1{\unskip}     \fi
\ifx \showLCCN     \undefined \def \showLCCN      #1{\unskip}     \fi
\ifx \shownote     \undefined \def \shownote      #1{#1}          \fi
\ifx \showarticletitle \undefined \def \showarticletitle #1{#1}   \fi
\ifx \showURL      \undefined \def \showURL       {\relax}        \fi
\providecommand\bibfield[2]{#2}
\providecommand\bibinfo[2]{#2}
\providecommand\natexlab[1]{#1}
\providecommand\showeprint[2][]{arXiv:#2}

\bibitem[Ahn et~al\mbox{.}(2022)]%
        {AhnCLGKJRPMK22}
\bibfield{author}{\bibinfo{person}{Minseon Ahn}, \bibinfo{person}{Andrew Chang}, \bibinfo{person}{Donghun Lee}, \bibinfo{person}{Jongmin Gim}, \bibinfo{person}{Jungmin Kim}, \bibinfo{person}{Jaemin Jung}, \bibinfo{person}{Oliver Rebholz}, \bibinfo{person}{Vincent Pham}, \bibinfo{person}{Krishna~T. Malladi}, {and} \bibinfo{person}{Yang{-}Seok Ki}.} \bibinfo{year}{2022}\natexlab{}.
\newblock \showarticletitle{Enabling {CXL} Memory Expansion for In-Memory Database Management Systems}. In \bibinfo{booktitle}{\emph{{DaMoN}}}. \bibinfo{publisher}{{ACM}}, \bibinfo{pages}{8:1--8:5}.
\newblock


\bibitem[Ahn et~al\mbox{.}(2024)]%
        {AhnWMLDBKSRR24}
\bibfield{author}{\bibinfo{person}{Minseon Ahn}, \bibinfo{person}{Thomas Willhalm}, \bibinfo{person}{Norman May}, \bibinfo{person}{Donghun Lee}, \bibinfo{person}{Suprasad~Mutalik Desai}, \bibinfo{person}{Daniel Booss}, \bibinfo{person}{Jungmin Kim}, \bibinfo{person}{Navneet Singh}, \bibinfo{person}{Daniel Ritter}, {and} \bibinfo{person}{Oliver Rebholz}.} \bibinfo{year}{2024}\natexlab{}.
\newblock \showarticletitle{An Examination of {CXL} Memory Use Cases for In-Memory Database Management Systems using {SAP} {HANA}}.
\newblock \bibinfo{journal}{\emph{Proc. {VLDB} Endow.}} \bibinfo{volume}{17}, \bibinfo{number}{12} (\bibinfo{year}{2024}), \bibinfo{pages}{3827--3840}.
\newblock
\href{https://doi.org/10.14778/3685800.3685809}{doi:\nolinkurl{10.14778/3685800.3685809}}


\bibitem[Ailamaki(2021)]%
        {natassa2021}
\bibfield{author}{\bibinfo{person}{Anastasia Ailamaki}.} \bibinfo{year}{2021}\natexlab{}.
\newblock \bibinfo{title}{Accelerated Data Management Systems Through Real-Time Specialization}.
\newblock \bibinfo{howpublished}{Keynote presented at MICRO}.
\newblock


\bibitem[Albutiu et~al\mbox{.}(2012)]%
        {AlbutiuKN12}
\bibfield{author}{\bibinfo{person}{Martina{-}Cezara Albutiu}, \bibinfo{person}{Alfons Kemper}, {and} \bibinfo{person}{Thomas Neumann}.} \bibinfo{year}{2012}\natexlab{}.
\newblock \showarticletitle{Massively Parallel Sort-Merge Joins in Main Memory Multi-Core Database Systems}.
\newblock \bibinfo{journal}{\emph{{VLDB}}} \bibinfo{volume}{5}, \bibinfo{number}{10} (\bibinfo{year}{2012}), \bibinfo{pages}{1064--1075}.
\newblock
\href{https://doi.org/10.14778/2336664.2336678}{doi:\nolinkurl{10.14778/2336664.2336678}}


\bibitem[Amazon(2024a)]%
        {amazon-ec2}
\bibfield{author}{\bibinfo{person}{Amazon}.} \bibinfo{year}{2024}\natexlab{a}.
\newblock \bibinfo{booktitle}{\emph{Amazon EC2 Instance types}}.
\newblock
\urldef\tempurl%
\url{https://aws.amazon.com/ec2/instance-types/}
\showURL{%
Retrieved June 24, 2024 from \tempurl}


\bibitem[Amazon(2024b)]%
        {amazon-graviton}
\bibfield{author}{\bibinfo{person}{Amazon}.} \bibinfo{year}{2024}\natexlab{b}.
\newblock \bibinfo{booktitle}{\emph{AWS Graviton Processors}}.
\newblock
\urldef\tempurl%
\url{https://aws.amazon.com/ec2/graviton/}
\showURL{%
Retrieved July 28, 2024 from \tempurl}


\bibitem[Amazon(2025a)]%
        {amazon-ec2-m5}
\bibfield{author}{\bibinfo{person}{Amazon}.} \bibinfo{year}{2025}\natexlab{a}.
\newblock \bibinfo{booktitle}{\emph{Amazon EC2 m5a.24xlarge instance}}.
\newblock
\urldef\tempurl%
\url{https://aws.amazon.com/ec2/instance-types/m5/}
\showURL{%
Retrieved May 8, 2025 from \tempurl}


\bibitem[Amazon(2025b)]%
        {amazon-ec2-x1}
\bibfield{author}{\bibinfo{person}{Amazon}.} \bibinfo{year}{2025}\natexlab{b}.
\newblock \bibinfo{booktitle}{\emph{Amazon EC2 x1.16xlarge instance}}.
\newblock
\urldef\tempurl%
\url{https://aws.amazon.com/ec2/instance-types/x1/}
\showURL{%
Retrieved May 8, 2025 from \tempurl}


\bibitem[Anandtech(2024)]%
        {intel-sierra-forest}
\bibfield{author}{\bibinfo{person}{Anandtech}.} \bibinfo{year}{2024}\natexlab{}.
\newblock \bibinfo{booktitle}{\emph{Intel Previews Sierra Forest with 288 E-Cores}}.
\newblock
\urldef\tempurl%
\url{https://www.anandtech.com/show/21276/intel-previews-sierra-forest-with-288-e-cores-announces-granite-rapids-d-for-2025-launch-at-mwc-2024}
\showURL{%
Retrieved July 28, 2024 from \tempurl}


\bibitem[Arulraj et~al\mbox{.}(2015)]%
        {ArulrajPD15}
\bibfield{author}{\bibinfo{person}{Joy Arulraj}, \bibinfo{person}{Andrew Pavlo}, {and} \bibinfo{person}{Subramanya Dulloor}.} \bibinfo{year}{2015}\natexlab{}.
\newblock \showarticletitle{Let's Talk About Storage {\&} Recovery Methods for Non-Volatile Memory Database Systems}. In \bibinfo{booktitle}{\emph{{SIGMOD}}}. \bibinfo{publisher}{{ACM}}, \bibinfo{pages}{707--722}.
\newblock


\bibitem[Balkesen et~al\mbox{.}(2013)]%
        {BalkesenATO13}
\bibfield{author}{\bibinfo{person}{Cagri Balkesen}, \bibinfo{person}{Gustavo Alonso}, \bibinfo{person}{Jens Teubner}, {and} \bibinfo{person}{M.~Tamer {\"{O}}zsu}.} \bibinfo{year}{2013}\natexlab{}.
\newblock \showarticletitle{Multi-Core, Main-Memory Joins: Sort vs. Hash Revisited}.
\newblock \bibinfo{journal}{\emph{{VLDB}}} \bibinfo{volume}{7}, \bibinfo{number}{1} (\bibinfo{year}{2013}), \bibinfo{pages}{85--96}.
\newblock
\href{https://doi.org/10.14778/2732219.2732227}{doi:\nolinkurl{10.14778/2732219.2732227}}


\bibitem[Bang et~al\mbox{.}(2020)]%
        {BangMPB20}
\bibfield{author}{\bibinfo{person}{Tiemo Bang}, \bibinfo{person}{Norman May}, \bibinfo{person}{Ilia Petrov}, {and} \bibinfo{person}{Carsten Binnig}.} \bibinfo{year}{2020}\natexlab{}.
\newblock \showarticletitle{The tale of 1000 Cores: an evaluation of concurrency control on real(ly) large multi-socket hardware}. In \bibinfo{booktitle}{\emph{{DaMoN}}}. \bibinfo{publisher}{{ACM}}, \bibinfo{pages}{3:1--3:9}.
\newblock
\href{https://doi.org/10.1145/3399666.3399910}{doi:\nolinkurl{10.1145/3399666.3399910}}


\bibitem[Bi et~al\mbox{.}(2024)]%
        {abs-2401-02954}
\bibfield{author}{\bibinfo{person}{Xiao Bi}, \bibinfo{person}{Deli Chen}, \bibinfo{person}{Guanting Chen}, \bibinfo{person}{Shanhuang Chen}, \bibinfo{person}{Damai Dai}, \bibinfo{person}{Chengqi Deng}, \bibinfo{person}{Honghui Ding}, \bibinfo{person}{Kai Dong}, \bibinfo{person}{Qiushi Du}, \bibinfo{person}{Zhe Fu}, \bibinfo{person}{Huazuo Gao}, \bibinfo{person}{Kaige Gao}, \bibinfo{person}{Wenjun Gao}, \bibinfo{person}{Ruiqi Ge}, \bibinfo{person}{Kang Guan}, \bibinfo{person}{Daya Guo}, \bibinfo{person}{Jianzhong Guo}, \bibinfo{person}{Guangbo Hao}, \bibinfo{person}{Zhewen Hao}, \bibinfo{person}{Ying He}, \bibinfo{person}{Wenjie Hu}, \bibinfo{person}{Panpan Huang}, \bibinfo{person}{Erhang Li}, \bibinfo{person}{Guowei Li}, \bibinfo{person}{Jiashi Li}, \bibinfo{person}{Yao Li}, \bibinfo{person}{Y.~K. Li}, \bibinfo{person}{Wenfeng Liang}, \bibinfo{person}{Fangyun Lin}, \bibinfo{person}{Alex~X. Liu}, \bibinfo{person}{Bo Liu}, \bibinfo{person}{Wen Liu}, \bibinfo{person}{Xiaodong Liu}, \bibinfo{person}{Xin Liu},
  \bibinfo{person}{Yiyuan Liu}, \bibinfo{person}{Haoyu Lu}, \bibinfo{person}{Shanghao Lu}, \bibinfo{person}{Fuli Luo}, \bibinfo{person}{Shirong Ma}, \bibinfo{person}{Xiaotao Nie}, \bibinfo{person}{Tian Pei}, \bibinfo{person}{Yishi Piao}, \bibinfo{person}{Junjie Qiu}, \bibinfo{person}{Hui Qu}, \bibinfo{person}{Tongzheng Ren}, \bibinfo{person}{Zehui Ren}, \bibinfo{person}{Chong Ruan}, \bibinfo{person}{Zhangli Sha}, \bibinfo{person}{Zhihong Shao}, \bibinfo{person}{Junxiao Song}, \bibinfo{person}{Xuecheng Su}, \bibinfo{person}{Jingxiang Sun}, \bibinfo{person}{Yaofeng Sun}, \bibinfo{person}{Minghui Tang}, \bibinfo{person}{Bingxuan Wang}, \bibinfo{person}{Peiyi Wang}, \bibinfo{person}{Shiyu Wang}, \bibinfo{person}{Yaohui Wang}, \bibinfo{person}{Yongji Wang}, \bibinfo{person}{Tong Wu}, \bibinfo{person}{Y. Wu}, \bibinfo{person}{Xin Xie}, \bibinfo{person}{Zhenda Xie}, \bibinfo{person}{Ziwei Xie}, \bibinfo{person}{Yiliang Xiong}, \bibinfo{person}{Hanwei Xu}, \bibinfo{person}{R.~X. Xu}, \bibinfo{person}{Yanhong Xu},
  \bibinfo{person}{Dejian Yang}, \bibinfo{person}{Yuxiang You}, \bibinfo{person}{Shuiping Yu}, \bibinfo{person}{Xingkai Yu}, \bibinfo{person}{B. Zhang}, \bibinfo{person}{Haowei Zhang}, \bibinfo{person}{Lecong Zhang}, \bibinfo{person}{Liyue Zhang}, \bibinfo{person}{Mingchuan Zhang}, \bibinfo{person}{Minghua Zhang}, \bibinfo{person}{Wentao Zhang}, \bibinfo{person}{Yichao Zhang}, \bibinfo{person}{Chenggang Zhao}, \bibinfo{person}{Yao Zhao}, \bibinfo{person}{Shangyan Zhou}, \bibinfo{person}{Shunfeng Zhou}, \bibinfo{person}{Qihao Zhu}, {and} \bibinfo{person}{Yuheng Zou}.} \bibinfo{year}{2024}\natexlab{}.
\newblock \showarticletitle{DeepSeek {LLM:} Scaling Open-Source Language Models with Longtermism}.
\newblock \bibinfo{journal}{\emph{CoRR}}  \bibinfo{volume}{abs/2401.02954} (\bibinfo{year}{2024}).
\newblock
\href{https://doi.org/10.48550/ARXIV.2401.02954}{doi:\nolinkurl{10.48550/ARXIV.2401.02954}}


\bibitem[Boeschen and Binnig(2022)]%
        {BoeschenB22}
\bibfield{author}{\bibinfo{person}{Nils Boeschen} {and} \bibinfo{person}{Carsten Binnig}.} \bibinfo{year}{2022}\natexlab{}.
\newblock \showarticletitle{GaccO - {A} GPU-accelerated {OLTP} {DBMS}}. In \bibinfo{booktitle}{\emph{{SIGMOD}}}. \bibinfo{publisher}{{ACM}}, \bibinfo{pages}{1003--1016}.
\newblock


\bibitem[Bre{\ss}(2013)]%
        {Bress13}
\bibfield{author}{\bibinfo{person}{Sebastian Bre{\ss}}.} \bibinfo{year}{2013}\natexlab{}.
\newblock \showarticletitle{Why it is time for a HyPE: {A} Hybrid Query Processing Engine for Efficient {GPU} Coprocessing in {DBMS}}.
\newblock \bibinfo{journal}{\emph{Proc. {VLDB} Endow.}} \bibinfo{volume}{6}, \bibinfo{number}{12} (\bibinfo{year}{2013}), \bibinfo{pages}{1398--1403}.
\newblock


\bibitem[Brown et~al\mbox{.}(2020)]%
        {BrownMRSKDNSSAA20}
\bibfield{author}{\bibinfo{person}{Tom~B. Brown}, \bibinfo{person}{Benjamin Mann}, \bibinfo{person}{Nick Ryder}, \bibinfo{person}{Melanie Subbiah}, \bibinfo{person}{Jared Kaplan}, \bibinfo{person}{Prafulla Dhariwal}, \bibinfo{person}{Arvind Neelakantan}, \bibinfo{person}{Pranav Shyam}, \bibinfo{person}{Girish Sastry}, \bibinfo{person}{Amanda Askell}, \bibinfo{person}{Sandhini Agarwal}, \bibinfo{person}{Ariel Herbert{-}Voss}, \bibinfo{person}{Gretchen Krueger}, \bibinfo{person}{Tom Henighan}, \bibinfo{person}{Rewon Child}, \bibinfo{person}{Aditya Ramesh}, \bibinfo{person}{Daniel~M. Ziegler}, \bibinfo{person}{Jeffrey Wu}, \bibinfo{person}{Clemens Winter}, \bibinfo{person}{Christopher Hesse}, \bibinfo{person}{Mark Chen}, \bibinfo{person}{Eric Sigler}, \bibinfo{person}{Mateusz Litwin}, \bibinfo{person}{Scott Gray}, \bibinfo{person}{Benjamin Chess}, \bibinfo{person}{Jack Clark}, \bibinfo{person}{Christopher Berner}, \bibinfo{person}{Sam McCandlish}, \bibinfo{person}{Alec Radford}, \bibinfo{person}{Ilya Sutskever},
  {and} \bibinfo{person}{Dario Amodei}.} \bibinfo{year}{2020}\natexlab{}.
\newblock \showarticletitle{Language Models are Few-Shot Learners}. In \bibinfo{booktitle}{\emph{{NeurIPS}}}.
\newblock
\urldef\tempurl%
\url{https://proceedings.neurips.cc/paper/2020/hash/1457c0d6bfcb4967418bfb8ac142f64a-Abstract.html}
\showURL{%
\tempurl}


\bibitem[Chen et~al\mbox{.}(2021)]%
        {ChenLRLGLASM21}
\bibfield{author}{\bibinfo{person}{Lili Chen}, \bibinfo{person}{Kevin Lu}, \bibinfo{person}{Aravind Rajeswaran}, \bibinfo{person}{Kimin Lee}, \bibinfo{person}{Aditya Grover}, \bibinfo{person}{Michael Laskin}, \bibinfo{person}{Pieter Abbeel}, \bibinfo{person}{Aravind Srinivas}, {and} \bibinfo{person}{Igor Mordatch}.} \bibinfo{year}{2021}\natexlab{}.
\newblock \showarticletitle{Decision Transformer: Reinforcement Learning via Sequence Modeling}. In \bibinfo{booktitle}{\emph{{NeurIPS}}}. \bibinfo{pages}{15084--15097}.
\newblock


\bibitem[Chrysogelos et~al\mbox{.}(2019)]%
        {ChrysogelosKAA19}
\bibfield{author}{\bibinfo{person}{Periklis Chrysogelos}, \bibinfo{person}{Manos Karpathiotakis}, \bibinfo{person}{Raja Appuswamy}, {and} \bibinfo{person}{Anastasia Ailamaki}.} \bibinfo{year}{2019}\natexlab{}.
\newblock \showarticletitle{HetExchange: Encapsulating heterogeneous {CPU-GPU} parallelism in {JIT} compiled engines}.
\newblock \bibinfo{journal}{\emph{Proc. {VLDB} Endow.}} \bibinfo{volume}{12}, \bibinfo{number}{5} (\bibinfo{year}{2019}), \bibinfo{pages}{544--556}.
\newblock


\bibitem[Cooper et~al\mbox{.}(2010)]%
        {CooperSTRS10}
\bibfield{author}{\bibinfo{person}{Brian~F. Cooper}, \bibinfo{person}{Adam Silberstein}, \bibinfo{person}{Erwin Tam}, \bibinfo{person}{Raghu Ramakrishnan}, {and} \bibinfo{person}{Russell Sears}.} \bibinfo{year}{2010}\natexlab{}.
\newblock \showarticletitle{Benchmarking cloud serving systems with {YCSB}}. In \bibinfo{booktitle}{\emph{{SoCC}}}. \bibinfo{publisher}{{ACM}}, \bibinfo{pages}{143--154}.
\newblock
\href{https://doi.org/10.1145/1807128.1807152}{doi:\nolinkurl{10.1145/1807128.1807152}}


\bibitem[Duplyakin et~al\mbox{.}(2019)]%
        {DuplyakinRMWDES19}
\bibfield{author}{\bibinfo{person}{Dmitry Duplyakin}, \bibinfo{person}{Robert Ricci}, \bibinfo{person}{Aleksander Maricq}, \bibinfo{person}{Gary Wong}, \bibinfo{person}{Jonathon Duerig}, \bibinfo{person}{Eric Eide}, \bibinfo{person}{Leigh Stoller}, \bibinfo{person}{Mike Hibler}, \bibinfo{person}{David Johnson}, \bibinfo{person}{Kirk Webb}, \bibinfo{person}{Aditya Akella}, \bibinfo{person}{Kuang{-}Ching Wang}, \bibinfo{person}{Glenn Ricart}, \bibinfo{person}{Larry Landweber}, \bibinfo{person}{Chip Elliott}, \bibinfo{person}{Michael Zink}, \bibinfo{person}{Emmanuel Cecchet}, \bibinfo{person}{Snigdhaswin Kar}, {and} \bibinfo{person}{Prabodh Mishra}.} \bibinfo{year}{2019}\natexlab{}.
\newblock \showarticletitle{The Design and Operation of CloudLab}. In \bibinfo{booktitle}{\emph{{USENIX}}}. \bibinfo{publisher}{{USENIX} Association}, \bibinfo{pages}{1--14}.
\newblock
\urldef\tempurl%
\url{https://www.usenix.org/conference/atc19/presentation/duplyakin}
\showURL{%
\tempurl}


\bibitem[Haas and Leis(2023)]%
        {HaasL23}
\bibfield{author}{\bibinfo{person}{Gabriel Haas} {and} \bibinfo{person}{Viktor Leis}.} \bibinfo{year}{2023}\natexlab{}.
\newblock \showarticletitle{What Modern NVMe Storage Can Do, And How To Exploit It: High-Performance {I/O} for High-Performance Storage Engines}.
\newblock \bibinfo{journal}{\emph{Proc. {VLDB} Endow.}} \bibinfo{volume}{16}, \bibinfo{number}{9} (\bibinfo{year}{2023}), \bibinfo{pages}{2090--2102}.
\newblock
\href{https://doi.org/10.14778/3598581.3598584}{doi:\nolinkurl{10.14778/3598581.3598584}}


\bibitem[He and Yu(2011)]%
        {HeY11}
\bibfield{author}{\bibinfo{person}{Bingsheng He} {and} \bibinfo{person}{Jeffrey~Xu Yu}.} \bibinfo{year}{2011}\natexlab{}.
\newblock \showarticletitle{High-throughput transaction executions on graphics processors}.
\newblock \bibinfo{journal}{\emph{Proc. {VLDB} Endow.}} \bibinfo{volume}{4}, \bibinfo{number}{5} (\bibinfo{year}{2011}), \bibinfo{pages}{314--325}.
\newblock


\bibitem[Izraelevitz et~al\mbox{.}(2019)]%
        {abs-1903-05714}
\bibfield{author}{\bibinfo{person}{Joseph Izraelevitz}, \bibinfo{person}{Jian Yang}, \bibinfo{person}{Lu Zhang}, \bibinfo{person}{Juno Kim}, \bibinfo{person}{Xiao Liu}, \bibinfo{person}{Amir~Saman Memaripour}, \bibinfo{person}{Yun~Joon Soh}, \bibinfo{person}{Zixuan Wang}, \bibinfo{person}{Yi Xu}, \bibinfo{person}{Subramanya~R. Dulloor}, \bibinfo{person}{Jishen Zhao}, {and} \bibinfo{person}{Steven Swanson}.} \bibinfo{year}{2019}\natexlab{}.
\newblock \showarticletitle{Basic Performance Measurements of the Intel Optane {DC} Persistent Memory Module}.
\newblock \bibinfo{journal}{\emph{CoRR}}  \bibinfo{volume}{abs/1903.05714} (\bibinfo{year}{2019}).
\newblock


\bibitem[Kraska et~al\mbox{.}(2019)]%
        {KraskaABCKLMMN19}
\bibfield{author}{\bibinfo{person}{Tim Kraska}, \bibinfo{person}{Mohammad Alizadeh}, \bibinfo{person}{Alex Beutel}, \bibinfo{person}{Ed~H. Chi}, \bibinfo{person}{Ani Kristo}, \bibinfo{person}{Guillaume Leclerc}, \bibinfo{person}{Samuel Madden}, \bibinfo{person}{Hongzi Mao}, {and} \bibinfo{person}{Vikram Nathan}.} \bibinfo{year}{2019}\natexlab{}.
\newblock \showarticletitle{SageDB: {A} Learned Database System}. In \bibinfo{booktitle}{\emph{{CIDR}}}. \bibinfo{publisher}{www.cidrdb.org}.
\newblock


\bibitem[Lambda(2025)]%
        {nvidia_gh_rent}
\bibfield{author}{\bibinfo{person}{Lambda}.} \bibinfo{year}{2025}\natexlab{}.
\newblock \bibinfo{booktitle}{\emph{Renting an NVIDIA Grace Hopper 2000 GPU}}.
\newblock
\urldef\tempurl%
\url{https://lambda.ai/nvidia-gh200}
\showURL{%
Retrieved May 8, 2025 from \tempurl}


\bibitem[Lang et~al\mbox{.}(2013)]%
        {LangLANK13}
\bibfield{author}{\bibinfo{person}{Harald Lang}, \bibinfo{person}{Viktor Leis}, \bibinfo{person}{Martina{-}Cezara Albutiu}, \bibinfo{person}{Thomas Neumann}, {and} \bibinfo{person}{Alfons Kemper}.} \bibinfo{year}{2013}\natexlab{}.
\newblock \showarticletitle{Massively Parallel NUMA-aware Hash Joins}. In \bibinfo{booktitle}{\emph{{IMDM}}}. \bibinfo{pages}{1--12}.
\newblock
\urldef\tempurl%
\url{http://www-db.in.tum.de/other/imdm2013/papers/Lang.pdf}
\showURL{%
\tempurl}


\bibitem[Lee et~al\mbox{.}(2024)]%
        {0001LBC24}
\bibfield{author}{\bibinfo{person}{Sangjin Lee}, \bibinfo{person}{Alberto Lerner}, \bibinfo{person}{Philippe Bonnet}, {and} \bibinfo{person}{Philippe Cudr{\'{e}}{-}Mauroux}.} \bibinfo{year}{2024}\natexlab{}.
\newblock \showarticletitle{Database Kernels: Seamless Integration of Database Systems and Fast Storage via {CXL}}. In \bibinfo{booktitle}{\emph{{CIDR}}}. \bibinfo{publisher}{www.cidrdb.org}.
\newblock


\bibitem[Leis et~al\mbox{.}(2014)]%
        {LeisBK014}
\bibfield{author}{\bibinfo{person}{Viktor Leis}, \bibinfo{person}{Peter~A. Boncz}, \bibinfo{person}{Alfons Kemper}, {and} \bibinfo{person}{Thomas Neumann}.} \bibinfo{year}{2014}\natexlab{}.
\newblock \showarticletitle{Morsel-driven parallelism: a NUMA-aware query evaluation framework for the many-core age}. In \bibinfo{booktitle}{\emph{{SIGMOD}}}. \bibinfo{publisher}{{ACM}}, \bibinfo{pages}{743--754}.
\newblock


\bibitem[Leis et~al\mbox{.}(2018)]%
        {LeisHK018}
\bibfield{author}{\bibinfo{person}{Viktor Leis}, \bibinfo{person}{Michael Haubenschild}, \bibinfo{person}{Alfons Kemper}, {and} \bibinfo{person}{Thomas Neumann}.} \bibinfo{year}{2018}\natexlab{}.
\newblock \showarticletitle{LeanStore: In-Memory Data Management beyond Main Memory}. In \bibinfo{booktitle}{\emph{{ICDE}}}. \bibinfo{publisher}{{IEEE} Computer Society}, \bibinfo{pages}{185--196}.
\newblock
\href{https://doi.org/10.1109/ICDE.2018.00026}{doi:\nolinkurl{10.1109/ICDE.2018.00026}}


\bibitem[Lerner and Alonso(2024)]%
        {LernerA24}
\bibfield{author}{\bibinfo{person}{Alberto Lerner} {and} \bibinfo{person}{Gustavo Alonso}.} \bibinfo{year}{2024}\natexlab{}.
\newblock \showarticletitle{{CXL} and the Return of Scale-Up Database Engines}.
\newblock \bibinfo{journal}{\emph{Proc. {VLDB} Endow.}} \bibinfo{volume}{17}, \bibinfo{number}{10} (\bibinfo{year}{2024}), \bibinfo{pages}{2568--2575}.
\newblock
\href{https://doi.org/10.14778/3675034.3675047}{doi:\nolinkurl{10.14778/3675034.3675047}}


\bibitem[Li et~al\mbox{.}(2013)]%
        {LiPMRL13}
\bibfield{author}{\bibinfo{person}{Yinan Li}, \bibinfo{person}{Ippokratis Pandis}, \bibinfo{person}{Ren{\'{e}} M{\"{u}}ller}, \bibinfo{person}{Vijayshankar Raman}, {and} \bibinfo{person}{Guy~M. Lohman}.} \bibinfo{year}{2013}\natexlab{}.
\newblock \showarticletitle{NUMA-aware algorithms: the case of data shuffling}. In \bibinfo{booktitle}{\emph{{CIDR}}}. \bibinfo{publisher}{www.cidrdb.org}.
\newblock


\bibitem[Liu et~al\mbox{.}(2024)]%
        {abs-2403-05821}
\bibfield{author}{\bibinfo{person}{Shu Liu}, \bibinfo{person}{Asim Biswal}, \bibinfo{person}{Audrey Cheng}, \bibinfo{person}{Xiangxi Mo}, \bibinfo{person}{Shiyi Cao}, \bibinfo{person}{Joseph~E. Gonzalez}, \bibinfo{person}{Ion Stoica}, {and} \bibinfo{person}{Matei Zaharia}.} \bibinfo{year}{2024}\natexlab{}.
\newblock \showarticletitle{Optimizing {LLM} Queries in Relational Workloads}.
\newblock \bibinfo{journal}{\emph{CoRR}}  \bibinfo{volume}{abs/2403.05821} (\bibinfo{year}{2024}).
\newblock
\href{https://doi.org/10.48550/ARXIV.2403.05821}{doi:\nolinkurl{10.48550/ARXIV.2403.05821}}
\showeprint[arXiv]{2403.05821}


\bibitem[Mageirakos et~al\mbox{.}(2022)]%
        {MageirakosMKCA22}
\bibfield{author}{\bibinfo{person}{Vasilis Mageirakos}, \bibinfo{person}{Riccardo Mancini}, \bibinfo{person}{Srinivas Karthik}, \bibinfo{person}{Bikash Chandra}, {and} \bibinfo{person}{Anastasia Ailamaki}.} \bibinfo{year}{2022}\natexlab{}.
\newblock \showarticletitle{Efficient GPU-accelerated Join Optimization for Complex Queries}. In \bibinfo{booktitle}{\emph{{ICDE}}}. \bibinfo{publisher}{{IEEE}}, \bibinfo{pages}{3190--3193}.
\newblock


\bibitem[NVIDIA(2024)]%
        {nvidia-cpu}
\bibfield{author}{\bibinfo{person}{NVIDIA}.} \bibinfo{year}{2024}\natexlab{}.
\newblock \bibinfo{booktitle}{\emph{NVIDIA Grace CPU Superchip Whitepaper}}.
\newblock
\urldef\tempurl%
\url{https://resources.nvidia.com/en-us-grace-cpu/nvidia-grace-cpu-superchip}
\showURL{%
Retrieved July 28, 2024 from \tempurl}


\bibitem[OpenAI(2023)]%
        {abs-2303-08774}
\bibfield{author}{\bibinfo{person}{OpenAI}.} \bibinfo{year}{2023}\natexlab{}.
\newblock \showarticletitle{{GPT-4} Technical Report}.
\newblock \bibinfo{journal}{\emph{CoRR}}  \bibinfo{volume}{abs/2303.08774} (\bibinfo{year}{2023}).
\newblock
\href{https://doi.org/10.48550/ARXIV.2303.08774}{doi:\nolinkurl{10.48550/ARXIV.2303.08774}}
\showeprint[arXiv]{2303.08774}


\bibitem[Pavlo et~al\mbox{.}(2017)]%
        {PavloAALLMMMPQS17}
\bibfield{author}{\bibinfo{person}{Andrew Pavlo}, \bibinfo{person}{Gustavo Angulo}, \bibinfo{person}{Joy Arulraj}, \bibinfo{person}{Haibin Lin}, \bibinfo{person}{Jiexi Lin}, \bibinfo{person}{Lin Ma}, \bibinfo{person}{Prashanth Menon}, \bibinfo{person}{Todd~C. Mowry}, \bibinfo{person}{Matthew Perron}, \bibinfo{person}{Ian Quah}, \bibinfo{person}{Siddharth Santurkar}, \bibinfo{person}{Anthony Tomasic}, \bibinfo{person}{Skye Toor}, \bibinfo{person}{Dana~Van Aken}, \bibinfo{person}{Ziqi Wang}, \bibinfo{person}{Yingjun Wu}, \bibinfo{person}{Ran Xian}, {and} \bibinfo{person}{Tieying Zhang}.} \bibinfo{year}{2017}\natexlab{}.
\newblock \showarticletitle{Self-Driving Database Management Systems}. In \bibinfo{booktitle}{\emph{{CIDR}}}. \bibinfo{publisher}{www.cidrdb.org}.
\newblock


\bibitem[Porobic et~al\mbox{.}(2014)]%
        {PorobicLTA14}
\bibfield{author}{\bibinfo{person}{Danica Porobic}, \bibinfo{person}{Erietta Liarou}, \bibinfo{person}{Pinar T{\"{o}}z{\"{u}}n}, {and} \bibinfo{person}{Anastasia Ailamaki}.} \bibinfo{year}{2014}\natexlab{}.
\newblock \showarticletitle{ATraPos: Adaptive transaction processing on hardware Islands}. In \bibinfo{booktitle}{\emph{{ICDE}}}. \bibinfo{publisher}{{IEEE} Computer Society}, \bibinfo{pages}{688--699}.
\newblock
\href{https://doi.org/10.1109/ICDE.2014.6816692}{doi:\nolinkurl{10.1109/ICDE.2014.6816692}}


\bibitem[Porobic et~al\mbox{.}(2012)]%
        {PorobicPBTA12}
\bibfield{author}{\bibinfo{person}{Danica Porobic}, \bibinfo{person}{Ippokratis Pandis}, \bibinfo{person}{Miguel Branco}, \bibinfo{person}{Pinar T{\"{o}}z{\"{u}}n}, {and} \bibinfo{person}{Anastasia Ailamaki}.} \bibinfo{year}{2012}\natexlab{}.
\newblock \showarticletitle{{OLTP} on Hardware Islands}.
\newblock \bibinfo{journal}{\emph{{VLDB}}} \bibinfo{volume}{5}, \bibinfo{number}{11} (\bibinfo{year}{2012}), \bibinfo{pages}{1447--1458}.
\newblock
\href{https://doi.org/10.14778/2350229.2350260}{doi:\nolinkurl{10.14778/2350229.2350260}}


\bibitem[Psaroudakis et~al\mbox{.}(2015)]%
        {PsaroudakisSMSA15}
\bibfield{author}{\bibinfo{person}{Iraklis Psaroudakis}, \bibinfo{person}{Tobias Scheuer}, \bibinfo{person}{Norman May}, \bibinfo{person}{Abdelkader Sellami}, {and} \bibinfo{person}{Anastasia Ailamaki}.} \bibinfo{year}{2015}\natexlab{}.
\newblock \showarticletitle{Scaling Up Concurrent Main-Memory Column-Store Scans: Towards Adaptive NUMA-aware Data and Task Placement}.
\newblock \bibinfo{journal}{\emph{{VLDB}}} \bibinfo{volume}{8}, \bibinfo{number}{12} (\bibinfo{year}{2015}), \bibinfo{pages}{1442--1453}.
\newblock


\bibitem[Psaroudakis et~al\mbox{.}(2016)]%
        {PsaroudakisSMSA16}
\bibfield{author}{\bibinfo{person}{Iraklis Psaroudakis}, \bibinfo{person}{Tobias Scheuer}, \bibinfo{person}{Norman May}, \bibinfo{person}{Abdelkader Sellami}, {and} \bibinfo{person}{Anastasia Ailamaki}.} \bibinfo{year}{2016}\natexlab{}.
\newblock \showarticletitle{Adaptive NUMA-aware data placement and task scheduling for analytical workloads in main-memory column-stores}.
\newblock \bibinfo{journal}{\emph{{VLDB}}} \bibinfo{volume}{10}, \bibinfo{number}{2} (\bibinfo{year}{2016}), \bibinfo{pages}{37--48}.
\newblock


\bibitem[Radford and Narasimhan(2018)]%
        {Radford2018ImprovingLU}
\bibfield{author}{\bibinfo{person}{Alec Radford} {and} \bibinfo{person}{Karthik Narasimhan}.} \bibinfo{year}{2018}\natexlab{}.
\newblock \showarticletitle{Improving Language Understanding by Generative Pre-Training}.
\newblock


\bibitem[Radford et~al\mbox{.}(2019)]%
        {Radford2019LanguageMA}
\bibfield{author}{\bibinfo{person}{Alec Radford}, \bibinfo{person}{Jeff Wu}, \bibinfo{person}{Rewon Child}, \bibinfo{person}{David Luan}, \bibinfo{person}{Dario Amodei}, {and} \bibinfo{person}{Ilya Sutskever}.} \bibinfo{year}{2019}\natexlab{}.
\newblock \showarticletitle{Language Models are Unsupervised Multitask Learners}.
\newblock


\bibitem[Schuh et~al\mbox{.}(2016)]%
        {SchuhCD16}
\bibfield{author}{\bibinfo{person}{Stefan Schuh}, \bibinfo{person}{Xiao Chen}, {and} \bibinfo{person}{Jens Dittrich}.} \bibinfo{year}{2016}\natexlab{}.
\newblock \showarticletitle{An Experimental Comparison of Thirteen Relational Equi-Joins in Main Memory}. In \bibinfo{booktitle}{\emph{{SIGMOD}}}. \bibinfo{publisher}{{ACM}}, \bibinfo{pages}{1961--1976}.
\newblock
\href{https://doi.org/10.1145/2882903.2882917}{doi:\nolinkurl{10.1145/2882903.2882917}}


\bibitem[Shah et~al\mbox{.}(2008)]%
        {ShahHWG08}
\bibfield{author}{\bibinfo{person}{Mehul~A. Shah}, \bibinfo{person}{Stavros Harizopoulos}, \bibinfo{person}{Janet~L. Wiener}, {and} \bibinfo{person}{Goetz Graefe}.} \bibinfo{year}{2008}\natexlab{}.
\newblock \showarticletitle{Fast scans and joins using flash drives}. In \bibinfo{booktitle}{\emph{{DaMoN}}}. \bibinfo{publisher}{{ACM}}, \bibinfo{pages}{17--24}.
\newblock


\bibitem[Sioulas et~al\mbox{.}(2019)]%
        {SioulasCKAA19}
\bibfield{author}{\bibinfo{person}{Panagiotis Sioulas}, \bibinfo{person}{Periklis Chrysogelos}, \bibinfo{person}{Manos Karpathiotakis}, \bibinfo{person}{Raja Appuswamy}, {and} \bibinfo{person}{Anastasia Ailamaki}.} \bibinfo{year}{2019}\natexlab{}.
\newblock \showarticletitle{Hardware-Conscious Hash-Joins on GPUs}. In \bibinfo{booktitle}{\emph{{ICDE}}}. \bibinfo{publisher}{{IEEE}}, \bibinfo{pages}{698--709}.
\newblock


\bibitem[Teubner and M{\"{u}}ller(2011)]%
        {TeubnerM11}
\bibfield{author}{\bibinfo{person}{Jens Teubner} {and} \bibinfo{person}{Ren{\'{e}} M{\"{u}}ller}.} \bibinfo{year}{2011}\natexlab{}.
\newblock \showarticletitle{How soccer players would do stream joins}. In \bibinfo{booktitle}{\emph{{SIGMOD}}}. \bibinfo{publisher}{{ACM}}, \bibinfo{pages}{625--636}.
\newblock
\href{https://doi.org/10.1145/1989323.1989389}{doi:\nolinkurl{10.1145/1989323.1989389}}


\bibitem[Touvron et~al\mbox{.}(2023)]%
        {abs-2302-13971}
\bibfield{author}{\bibinfo{person}{Hugo Touvron}, \bibinfo{person}{Thibaut Lavril}, \bibinfo{person}{Gautier Izacard}, \bibinfo{person}{Xavier Martinet}, \bibinfo{person}{Marie{-}Anne Lachaux}, \bibinfo{person}{Timoth{\'{e}}e Lacroix}, \bibinfo{person}{Baptiste Rozi{\`{e}}re}, \bibinfo{person}{Naman Goyal}, \bibinfo{person}{Eric Hambro}, \bibinfo{person}{Faisal Azhar}, \bibinfo{person}{Aur{\'{e}}lien Rodriguez}, \bibinfo{person}{Armand Joulin}, \bibinfo{person}{Edouard Grave}, {and} \bibinfo{person}{Guillaume Lample}.} \bibinfo{year}{2023}\natexlab{}.
\newblock \showarticletitle{LLaMA: Open and Efficient Foundation Language Models}.
\newblock \bibinfo{journal}{\emph{CoRR}}  \bibinfo{volume}{abs/2302.13971} (\bibinfo{year}{2023}).
\newblock
\urldef\tempurl%
\url{https://doi.org/10.48550/arXiv.2302.13971}
\showURL{%
\tempurl}


\bibitem[Tsirogiannis et~al\mbox{.}(2009)]%
        {TsirogiannisHSWG09}
\bibfield{author}{\bibinfo{person}{Dimitris Tsirogiannis}, \bibinfo{person}{Stavros Harizopoulos}, \bibinfo{person}{Mehul~A. Shah}, \bibinfo{person}{Janet~L. Wiener}, {and} \bibinfo{person}{Goetz Graefe}.} \bibinfo{year}{2009}\natexlab{}.
\newblock \showarticletitle{Query processing techniques for solid state drives}. In \bibinfo{booktitle}{\emph{{SIGMOD}}}. \bibinfo{publisher}{{ACM}}, \bibinfo{pages}{59--72}.
\newblock


\bibitem[van Renen et~al\mbox{.}(2018)]%
        {RenenLK0HODHS18}
\bibfield{author}{\bibinfo{person}{Alexander van Renen}, \bibinfo{person}{Viktor Leis}, \bibinfo{person}{Alfons Kemper}, \bibinfo{person}{Thomas Neumann}, \bibinfo{person}{Takushi Hashida}, \bibinfo{person}{Kazuichi Oe}, \bibinfo{person}{Yoshiyasu Doi}, \bibinfo{person}{Lilian Harada}, {and} \bibinfo{person}{Mitsuru Sato}.} \bibinfo{year}{2018}\natexlab{}.
\newblock \showarticletitle{Managing Non-Volatile Memory in Database Systems}. In \bibinfo{booktitle}{\emph{{SIGMOD}}}. \bibinfo{publisher}{{ACM}}, \bibinfo{pages}{1541--1555}.
\newblock


\bibitem[Vaswani et~al\mbox{.}(2017)]%
        {VaswaniSPUJGKP17}
\bibfield{author}{\bibinfo{person}{Ashish Vaswani}, \bibinfo{person}{Noam Shazeer}, \bibinfo{person}{Niki Parmar}, \bibinfo{person}{Jakob Uszkoreit}, \bibinfo{person}{Llion Jones}, \bibinfo{person}{Aidan~N. Gomez}, \bibinfo{person}{Lukasz Kaiser}, {and} \bibinfo{person}{Illia Polosukhin}.} \bibinfo{year}{2017}\natexlab{}.
\newblock \showarticletitle{Attention is All you Need}. In \bibinfo{booktitle}{\emph{{NeurIPS}}}. \bibinfo{pages}{5998--6008}.
\newblock


\bibitem[Wagner et~al\mbox{.}(2021)]%
        {WagnerK021}
\bibfield{author}{\bibinfo{person}{Benjamin Wagner}, \bibinfo{person}{Andr{\'{e}} Kohn}, {and} \bibinfo{person}{Thomas Neumann}.} \bibinfo{year}{2021}\natexlab{}.
\newblock \showarticletitle{Self-Tuning Query Scheduling for Analytical Workloads}. In \bibinfo{booktitle}{\emph{{SIGMOD}}}. \bibinfo{publisher}{{ACM}}, \bibinfo{pages}{1879--1891}.
\newblock


\bibitem[Wang(2017)]%
        {btreecode}
\bibfield{author}{\bibinfo{person}{Ziqi Wang}.} \bibinfo{year}{2017}\natexlab{}.
\newblock \bibinfo{booktitle}{\emph{Index Micro Benchmarks}}.
\newblock
\urldef\tempurl%
\url{https://github.com/wangziqi2016/index-microbench}
\showURL{%
Retrieved October 11, 2024 from \tempurl}


\bibitem[Wang et~al\mbox{.}(2018)]%
        {WangPLLZKA18}
\bibfield{author}{\bibinfo{person}{Ziqi Wang}, \bibinfo{person}{Andrew Pavlo}, \bibinfo{person}{Hyeontaek Lim}, \bibinfo{person}{Viktor Leis}, \bibinfo{person}{Huanchen Zhang}, \bibinfo{person}{Michael Kaminsky}, {and} \bibinfo{person}{David~G. Andersen}.} \bibinfo{year}{2018}\natexlab{}.
\newblock \showarticletitle{Building a Bw-Tree Takes More Than Just Buzz Words}. In \bibinfo{booktitle}{\emph{{SIGMOD}}}. \bibinfo{publisher}{{ACM}}, \bibinfo{pages}{473--488}.
\newblock
\href{https://doi.org/10.1145/3183713.3196895}{doi:\nolinkurl{10.1145/3183713.3196895}}


\bibitem[Wikipedia(2024)]%
        {apple-silicon}
\bibfield{author}{\bibinfo{person}{Wikipedia}.} \bibinfo{year}{2024}\natexlab{}.
\newblock \bibinfo{booktitle}{\emph{Apple Silicon Processors}}.
\newblock
\urldef\tempurl%
\url{https://en.wikipedia.org/wiki/Apple_silicon}
\showURL{%
Retrieved July 28, 2024 from \tempurl}


\bibitem[Williams and Zipser(1989)]%
        {WilliamsZ89}
\bibfield{author}{\bibinfo{person}{Ronald~J. Williams} {and} \bibinfo{person}{David Zipser}.} \bibinfo{year}{1989}\natexlab{}.
\newblock \showarticletitle{A Learning Algorithm for Continually Running Fully Recurrent Neural Networks}.
\newblock \bibinfo{journal}{\emph{Neural Comput.}} \bibinfo{volume}{1}, \bibinfo{number}{2} (\bibinfo{year}{1989}), \bibinfo{pages}{270--280}.
\newblock
\href{https://doi.org/10.1162/NECO.1989.1.2.270}{doi:\nolinkurl{10.1162/NECO.1989.1.2.270}}


\bibitem[Zhou et~al\mbox{.}(2021)]%
        {ZhouAPC21}
\bibfield{author}{\bibinfo{person}{Xinjing Zhou}, \bibinfo{person}{Joy Arulraj}, \bibinfo{person}{Andrew Pavlo}, {and} \bibinfo{person}{David~E. Cohen}.} \bibinfo{year}{2021}\natexlab{}.
\newblock \showarticletitle{Spitfire: {A} Three-Tier Buffer Manager for Volatile and Non-Volatile Memory}. In \bibinfo{booktitle}{\emph{{SIGMOD}}}. \bibinfo{publisher}{{ACM}}, \bibinfo{pages}{2195--2207}.
\newblock


\bibitem[Zuo et~al\mbox{.}(2018)]%
        {Zuo0W18}
\bibfield{author}{\bibinfo{person}{Pengfei Zuo}, \bibinfo{person}{Yu Hua}, {and} \bibinfo{person}{Jie Wu}.} \bibinfo{year}{2018}\natexlab{}.
\newblock \showarticletitle{Write-Optimized and High-Performance Hashing Index Scheme for Persistent Memory}. In \bibinfo{booktitle}{\emph{{OSDI}}}. \bibinfo{publisher}{{USENIX} Association}, \bibinfo{pages}{461--476}.
\newblock


\end{thebibliography}


\end{document}